\documentclass{emulateapj}
\usepackage{graphicx,amssymb,amsmath,times}

\begin{document}
%Title of paper
\title{MAGIC observations of very high energy $\gamma$-rays from HESS J1813-178}

\author{
 J.~Albert\altaffilmark{a}, 
 E.~Aliu\altaffilmark{b}, 
 H.~Anderhub\altaffilmark{c}, 
 P.~Antoranz\altaffilmark{d}, 
 A.~Armada\altaffilmark{b}, 
 M.~Asensio\altaffilmark{d}, 
 C.~Baixeras\altaffilmark{e}, 
 J.~A.~Barrio\altaffilmark{d}, 
 M.~Bartelt\altaffilmark{f}, 
 H.~Bartko\altaffilmark{g,*}
% ...
}
 \altaffiltext{a} {Universit\"at W\"urzburg, D-97074 W\"urzburg, Germany}
 \altaffiltext{b} {Institut de F\'\i sica d'Altes Energies, Edifici Cn., E-08193 Bellaterra (Barcelona), Spain}
 \altaffiltext{c} {ETH Z\"urich, CH-8093 H\"onggerberg, Switzerland}
 \altaffiltext{d} {Universidad Complutense, E-28040 Madrid, Spain}
 \altaffiltext{e} {Universitat Aut\`onoma de Barcelona, E-08193 Bellaterra, Spain}
 \altaffiltext{f} {Universit\"at Dortmund, D-44227 Dortmund, Germany}
 \altaffiltext{g} {Max-Planck-Institut f\"ur Physik, D-80805 M\"unchen, Germany}
% ...
\altaffiltext{*} {correspondence: hbartko@mppmu.mpg.de}

%% abstract %%%%%%%%%%%%%%%%%%%%%%%%%%%%%%%%%%%%%%%%%%%%%%%%%

\begin{abstract}

Recently, the HESS collaboration has reported the detection of
$\gamma$-ray emission above a few hundred GeV from eight new
sources located close to the Galactic Plane. The source HESS
J1813-178 has sparked particular interest, as subsequent radio
observations imply an association with SNR G12.82-0.02. Triggered
by the detection in VHE $\gamma$-rays, a positionally coincident 
source has also been
found in INTEGRAL and ASCA data. In this {\it Letter} we present
MAGIC observations of HESS J1813-178, resulting in the detection
of a differential $\gamma$-ray flux consistent with a hard-slope
power law, described as $\mathrm{d}N_{\gamma}/(\mathrm{d}A
\mathrm{d}t \mathrm{d}E) = (3.3 \pm 0.5) \times 10^{-12}
(E/\mathrm{TeV})^{-2.1 \pm 0.2} \ \mathrm{cm}^{-2}\mathrm{s}^{-1}
\mathrm{TeV}^{-1}$. We briefly discuss the observational technique
used, the procedure implemented for the data analysis, and put
this detection in the perspective of multifrequency observations.

\end{abstract}


\section{Introduction}

In the Galactic Plane scan performed with the HESS Cherenkov array in 2004,
with a flux sensitivity of $3\%$ Crab units for
$\gamma$-rays above $100$ GeV, eight sources
were discovered \citep{Aharonian2005a,Aharonian2005b}. One of the newly detected
$\gamma$-ray sources was HESS J1813-178. At the beginning, HESS J1813-178 could not be 
identified and was assumed to be a ``dark particle accelerator'', without
reported counterparts at lower frequencies.
% (referred to as HESS J1813-178 in the rest of this {\it Letter})

Since the original discovery, HESS J1813-178 has been associated with
the supernova remnant
SNR G12.82-0.02 \citep{Ubertini2005,Brogan2005,Helfand2005}.
One may still not exclude this coincidence being the
result of just a chance association. \citet{Aharonian2005a} state a probability of (6\%) that one of their new sources is by chance spatially consistent with a SNR. Nevertheless, the properties of SNR G12.82-0.02, the
multifrequency spectral energy distribution, and the flux and
spectrum of the high energy $\gamma$-rays detected from this
direction appear to be consistent with a SNR origin.  
%, the probability of which is
%non-negligible (6\%) in the central region of the Galaxy
%\citep{Aharonian2005a}.

HESS J1813-178 has been found to be nearly point like ($\sigma_{\mathrm{source}}=2.2'$) by \citet{Aharonian2005b}. Given the size of the
SNR, the angular resolution of the HESS telescope, and the depth of the
observations, the source size does not rule out a possible shell
origin. The $\gamma$-ray source lies at $10'$  from the center
of the radio source W33. This patch of the sky is highly obscured
and has indications of being a recent star formation region
\citep{Hess27}.

In this {\it Letter} we present Major Atmospheric Gamma Imaging
Cherenkov telescope (MAGIC) observations of HESS J1813-178. We
briefly discuss the observational technique used and the procedure
implemented for the data analysis. We derive a high energy $\gamma$-ray
spectrum of the source, and analyze it in the perspective of
multifrequency observations.

\newpage

\section{Observations}

\begin{figure}
\begin{center}
\includegraphics[totalheight=7cm]{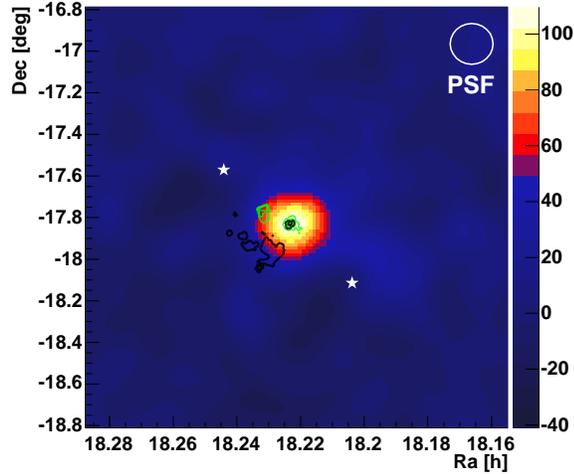}
\end{center}
\caption{ Sky map of $\gamma$-ray candidate events (background subtracted) in the
directions of HESS J1813-178 for SIZE $\geq 600$ ph. el.
(corresponding to an energy threshold of about 1~TeV). Overlayed are contours of 90 cm VLA radio (black) and ASCA X-ray data (green) from \citet{Brogan2005}. The two white stars denote the tracking positions W1,W2 in the wobble mode.}
\label{fig:disp_map}
\end{figure}

MAGIC (see e.g., \citet{MAGIC-commissioning,CortinaICRC}) is
currently the largest single dish Imaging
Air Cherenkov Telescope (IACT) in operation. Located on the Canary
Island La Palma ($28.8^\circ$N, $17.8^\circ$W, 2200~m a.s.l), the
telescope has a 17-m diameter tessellated parabolic mirror,
supported by a light weight carbon fiber frame. It is equipped
with a high efficiency 576-pixel $3.5^\circ$ field-of-view photomultiplier
camera. The analogue signals are transported via optical fibers to
the trigger electronics
and are read out by a 300 MSamples/s FADC system.%the 300 MHz FADC readout.

At La Palma, HESS J1813-178 culminates at about $47^\circ$ zenith angle (ZA). The
large ZA implies a high energy threshold of about 400~GeV for MAGIC
observations. It also provides
a large effective collection area (see e.g. \citet{LZA-sensitivity}).
The sky region around the location of HESS J1813-178 has a relatively high
and non-uniform level of light.
Within a distance of 1$^{\circ}$ from HESS J1813-178, there are no
stars brighter than 8$^{\mathrm{th}}$ magnitude, with the star field being
brighter in the region SW of the source.

The MAGIC observations were carried out in the false-source
tracking (wobble) mode \citep{wobble}. The sky directions (W1, W2)
to be tracked are chosen such that in the camera the sky field relative
to the source position is similar to the sky field relative to the mirror 
source position (anti-source position). The
source direction is in both cases $0.4^\circ$ offset from the
camera center. In Fig. \ref{fig:disp_map} these two tracking
positions are shown by white stars. During wobble mode data
taking, 50\% of the data is taken at W1 and 50\% at W2, switching
(wobbling) between the 2 directions every 30 minutes. This
observation mode allowed a reliable background estimation least
affected by the large ZA and inhomogeneous star field.

\section{Data Analysis}

\begin{figure}
\begin{center}
\includegraphics[totalheight=5cm]{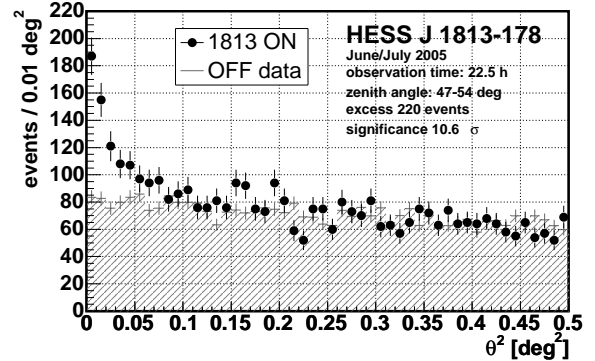}
\end{center}
\caption{ Distributions of $\theta^2$ values for the source and
anti-source, see text, for SIZE $\geq 600$ ph. el.
(corresponding to an energy threshold of about 1~TeV).}
\label{fig:theta2}
\end{figure}

HESS J1813-178 was observed for a total of 25 hours in the period
June-July 2005  (ZA $\leq 52^\circ$). About 15 million
triggers have been recorded. The calibration of the raw data of
the MAGIC camera is explained in \citet{MAGIC_calibration}. Image
cleaning tail cuts were applied: Pixels are only considered to be
part of the image if their reconstructed charge signal is larger
than 10 photo electrons (ph.el.) (core pixels) or if their charge is larger
than 5 ph. el.
and they have at least one neighboring core pixel. These tail cuts are accordingly scaled for the larger outer pixels of the MAGIC camera. The camera images
are characterized by image parameters \citep{Hillas_parameters}. After the image cleaning and rejection of broken runs about 10 million events remained for further analysis. These data were processed for $\gamma$/hadron
separation, in a similar way as described in \citet{Fegan1997}.

\begin{figure}
\begin{center}
\includegraphics[totalheight=5cm]{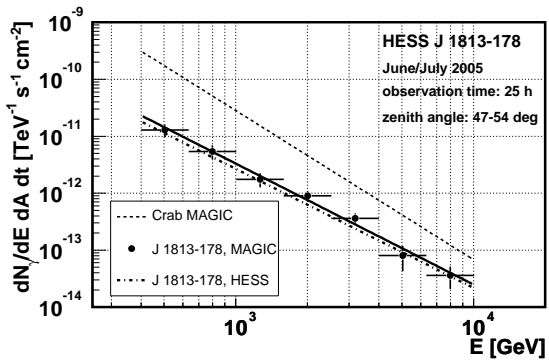}
\end{center}
\caption{ Reconstructed VHE $\gamma$-ray spectrum of HESS J1813-178.
The spectral index is $-2.1\pm0.2$ and the integral flux above
400~GeV is about 8\% of the Crab nebula (statistical errors only).
The dashed line shows the spectrum of the Crab nebula as measured
by MAGIC \citep{Crab_MAGIC}. The dot-dashed line shows the results of the HESS collaboration \citep{Aharonian2005b}.}
\label{fig:prelspectrum}
\end{figure}

In this analysis, the Random Forest method (see \citet{RF} for a
detailed description) was applied for the $\gamma$/hadron
separation and the energy estimation. For the training of the
Random Forest a sample of Monte Carlo (MC) generated
$\gamma$-showers was used to represent the $\gamma$-rays and a
randomly chosen sub-set of the measured data was used to represent
the background. The MC
$\gamma$-showers were generated between 47$^\circ$ and 54$^\circ$
ZA with energies between 10~GeV and 30~TeV. The spectral index of
the generated differential spectrum $dN_{\gamma}/dE \sim E^{\Gamma}$
was chosen as $\Gamma=-2.6$, in
agreement with the MAGIC-observed energy spectrum of the Crab
nebula \citep{Crab_MAGIC}.\footnote{In order to develop and verify
the analysis at high zenith angles, Crab data in the interesting
ZA range around 50$^{\circ}$ were taken in January 2005. From that
sample, we determined the Crab energy spectrum and found it to be
consistent with other existing measurements (see Fig.
\ref{fig:prelspectrum}, upper curve).} The source-position independent image parameters  SIZE, WIDTH, LENGTH, CONC \citep{Hillas_parameters} and the third moment along the major image axis, as well as the source-position dependent parameter DIST \citep{Hillas_parameters}, were
selected to parameterize the shower images. After the training,
the Random Forest method allows to calculate for every event a
parameter, dubbed hadronness, which is a measure of the
probability that the event belongs to the background. The
$\gamma$-sample is defined by selecting showers with a hadronness
below a specified value. An independent sample of MC $\gamma$-showers was used to determine the efficiency of the applied cuts.

For each event, its original sky position is determined by using
the DISP-method \citep{wobble,Lessard2001}. At this stage only source independent image parameters are used in the RF training. Figure
\ref{fig:disp_map} shows the sky map of $\gamma$-ray candidate
events (background subtracted) from the direction of HESS J1813-178. It is smoothed with a two-dimensional Gaussian with a standard deviation of $0.1^{\circ}$. To provide a good angular resolution a tight hadronness cut and a lower SIZE cut of
600 ph. el. has been applied. The SIZE cut corresponds to an energy
threshold of about 1~TeV. The sky map is overlayed with contours of 90
cm VLA radio and ASCA X-ray data from \citet{Brogan2005}. The
excess is centered at (RA,
DEC)=(18$^\mathrm{h}$13$^\mathrm{m}$27$^\mathrm{s}$,
-17$^\circ 48' 40''$) and coincides well with the position of SNR
G12.82-0.02. The systematic pointing uncertainty is estimated to
be $2'$ (for description of the MAGIC telescope drive system see \citet{Bretz2003}) and might in future be greatly reduced with the MAGIC starfield
monitor \citep{starguider}. Apart from the main excess coincident with HESS J1813-178 there are no other significant excesses present.

Figure \ref{fig:theta2} shows the distribution of the squared
angular distance, $\theta^2$, between the reconstructed shower
direction and the nominal object position. The observed excess in
the direction of HESS J1813-178 has a significance of 10.6
standard deviations (according to equation 17 of \citet{Li_Ma1983}). Within errors the source position and the flux level
are compatible with the
measurement of HESS \citep{Aharonian2005b}.

For the determination of the energy spectrum, the RF was trained including the source dependent image parameter DIST with respect to the nominal excess position. A loose cut on the hadronness was used. Above the low energy turn-on, the cut efficiency reaches about 70\% corresponding to an effective collection area for $\gamma$-ray showers of about 180000 m$^2$.
Figure \ref{fig:prelspectrum} shows the reconstructed very high
energy $\gamma$-ray spectrum of HESS J1813-178 after the unfolding with
the instrumental energy resolution. The measured spectral points are fit by a simple power law spectrum, taking the full instrumental energy resolution into account as described in \citet{Mizobuchi2005}. The result is given by ($\chi^2/\mathrm{n.d.f}=5.3/5$):
\begin{eqnarray*} \frac{ \mathrm{d}N_{\gamma}}{\mathrm{d}A \mathrm{d}t \mathrm{d}E} =
(3.3 \pm 0.5) \times  10^{-12} (E/\mathrm{TeV})^{-2.1 \pm 0.2}
\nonumber
\\
\hspace{4cm} \mathrm{cm}^{-2}\mathrm{s}^{-1}
\mathrm{TeV}^{-1}.\end{eqnarray*} The quoted errors ($1\sigma$) are purely
statistical. The systematic error is estimated to be 35\%
in the flux level determination and 0.2 in the spectral index.
Within errors the flux is
steady in the time scales explored within these observations
(weeks), as well as in the year-long time-span between the MAGIC
and HESS pointings.

\section{Discussion}

\begin{figure}
\begin{center}
\includegraphics[height=7cm]{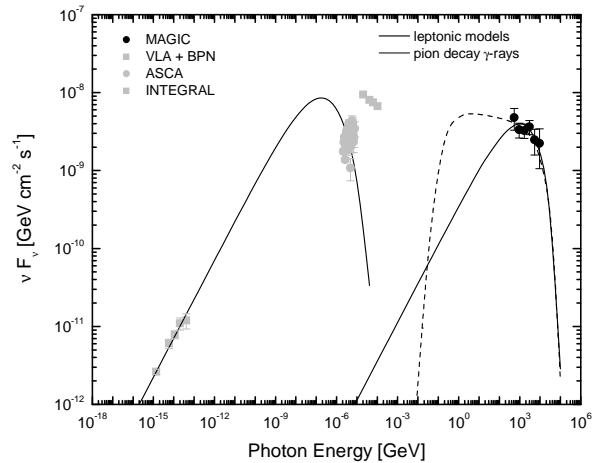}
\end{center}
\caption{Leptonic and hadronic models for the J1813-178 data.
Details are given in the text. Radio data is from VLA, Bonn,
Parkes, and Nobeyama observatories (Brogan et al. 2005), X-ray and
hard X-ray data are from ASCA and INTEGRAL (Ubertini et al. 2005,
Brogan et al. 2005).} \label{fig:modelling}
\end{figure}

Shortly after the discovery of HESS J1813-178, X-ray emission was found
in ASCA data coming from the source AX J1813-178 (a.k.a.
AGPS~273.4-17.8), \citet{Brogan2005}, also \citet{Ubertini2005}.
X-ray emission detected by ASCA is predominantly non-thermal, and
compatible with that expected from a PWN or a SNR shell. X-ray pulsed
emission has not been detected, but the quality and amount of the
data does not imply strong constraints \citep{Brogan2005}.
Statistically, X-ray data are not good enough to unambiguously
separate a pure power law from power law + thermal contribution
either. However, in both cases, none of the two component fits
yield a significantly different absorbing column density or
photon index when compared with a single power law fit
\citep{Brogan2005}.  Data analysis from the INTEGRAL satellite
also showed a soft and luminous source at the same location, in
the 20-100 keV range \citep{Ubertini2005}.

Simultaneously with this X-ray match, HESS J1813-178 was also found as
a non-thermal source in radio data, using observations with VLA
(90 and 20 cm), Bonn (3 cm), Parkes (6 cm), and Nobeyama (3 cm)
telescopes \citep{Brogan2005,Helfand2005}. These groups discovered
a shell-type supernova remnant (SNR G12.8-0.0) with a section of
the shell coinciding with HESS J1813-178. The radio spectral index
was found to be -0.48 $\pm$ 0.03. There are no known radio pulsars
detected at the HESS J1813-178 position \citep{Manchester2005}. 

\citet{Brogan2005} conclude that SNR G12.8-0.0 should 
lie at or beyond the distance of W33, $\sim4$ kpc. They have derived
a high column density of $N_H \sim 10^{23}$ cm$^{-2}$ from the ASCA data
which suggests a significant source of absorption in the foreground.

The multi-wavelength emission coming from the direction of HESS
J1813-178 is shown in Fig. \ref{fig:modelling}, including the
new MAGIC data at high energies. We have compared hadronic
(neutral pion decay) and leptonic (inverse Compton) emission
models with the high energy $\gamma$-ray data, for a review see \citet{Torres2002}. In the case of
hadronic models, the observed $\gamma$-ray luminosity ($2.5
\times 10^{34}$ erg s$^{-1}$ between 0.4 - 6 TeV, at 4 kpc)
implies that the required density of matter is $\sim 6$ cm$^{-3}$. The
$\gamma$-ray production region is presumed to be the whole SNR volume. An acceleration efficiency of hadrons of the
order of 3\% and a supernova power of $10^{51}$ erg was assumed. 
Therefore,
relativistic hadrons need only about 2 $M_\odot$ of target mass
inside the SNR volume ($\sim$1.5 pc radius at 4 kpc distance) in
order to be consistent with the observed luminosity.
Alternatively, the target mass for the cosmic ray spectrum can be
provided by a small cloud of a few solar masses located in a
region close to the SNR shell. For the model shown in Fig.
\ref{fig:modelling}, we have assumed that the proton distribution
is described by $\mathrm{d}N_{p}/(\mathrm{d}V
\mathrm{d}E)=A_p (E/{\rm GeV})^{-\alpha} \exp{(-E/E_{\rm max})}\;
{\rm GeV^{-1} cm^{-3}} $ where $A_p$ is a dimensionless
normalization constant. We found that $\alpha=2.1$ and $E_{\rm
max}=100$ TeV is a good fit to the data, and that the
normalization constant is such that the amount of supernova
explosion energy converted into relativistic cosmic rays needs not
to be more than a few percent to agree with MAGIC observations.

For leptonic models, we have assumed a similarly described
distribution of relativistic electrons
$\mathrm{d}N_{e}/(\mathrm{d}V \mathrm{d}E)=A_e (E/{\rm
GeV})^{-\alpha_e} \exp{(-E/E_{\rm max, e})}\; {\rm GeV^{-1}
cm^{-3}} $. We found that several different inverse Compton models
produce reasonable good fits at high energies, e.g. $\alpha_e=2.0$
and $E_{\rm max, e} = 20$ TeV; and $\alpha_e=2.1$ and 2.2 and
$E_{\rm max, e} = 30$ TeV, all having their energetics at ease
with the energy assumed to be released by the supernova explosion. 
The source of target photons for these models was assumed to be the
cosmic microwave background.
The radio spectrum at lower energies is fitted 
best by a slope of $\alpha_e=2.0$. It is computed as
synchrotron emission of the same electron population. 
In the model
shown in Fig. \ref{fig:modelling}, a magnetic filling fraction
of about 20\%, a magnetic field of 10 $\mu$G and $E_{\rm max, e} = 20$ TeV 
and $\alpha_e=2.0$ have been adopted.
This model is similar to one of the models presented by \citet{Brogan2005}
(blue line in their Fig. 3, $E_{\rm max, e} = 30$, $\alpha_e=2.0$),
without yet having a high energy $\gamma$-ray spectrum.
The soft X-ray data (ASCA) plotted correspond to those shown by
Brogan et al. (2005) (black data points in their Fig. 3). 
Our model, as well as Brogan et~al.'s, is in rough agreement with MAGIC data,
radio and roughly also with the X-ray data, even if the spectral behaviour in the
ASCA energy range (the slope) is somewhat different. That is, we have
priviledged the radio data in fitting this SED as an example, a better
X-ray fitting could be obtained at the expense of a worse radio one.
It is worth noticing that the hard
X-ray data of INTEGRAL \citep{Ubertini2005} cannot be fitted well
with the same electron population that is assumed to produce
higher and lower energy photons. Nevertheless, as one
can see in Fig. 2 of \citep{Ubertini2005}, the ASCA, MAGIC and HESS sources match spatially well, while the INTEGRAL/IBIS source is spatially only marginally compatible with them.
Further multi-frequency observations in the
x-ray and hard x-ray regime are needed to obtain definite
conclusions up to what level the leptonic models need two
electron populations.

\section{Concluding Remarks}

The detection of HESS J1813-178 using the MAGIC Telescope confirms a
new very high-energy $\gamma$-ray source in the Galactic Plane. A
reasonably large data set was collected from observations at large
zenith angles to infer the spectrum of this source up to energies
of about 10~TeV.  Between 400~GeV and 10~TeV the differential energy
spectrum can be fitted with a power law of slope $\Gamma=-2.1\pm
0.2$.  These data can be used to cross-calibrate HESS and MAGIC,
their independent observations show satisfactory agreement.

Multifrequency data in the radio, X-ray, and $\gamma$-ray band
imply a connection between HESS J1813-178 and SNR G12.82-0.02
\citep{Helfand2005,Ubertini2005,Brogan2005}. Generally, hard
$\gamma$-ray spectra are expected from SNRs due to Fermi
acceleration of cosmic rays \citep{Ginzburg1964,Torres2002}.
The hard spectrum determined for HESS J1813-178 may be a further hint
for its association with the SNR G12.82-0.02.

Present data are not sufficient to discriminate between existing models for different acceleration mechanisms. Future observations at lower energies with improved gamma-ray telescopes and/or the GLAST satellite will undoubtedly permit to shed more light on the existing leptonic and hadronic models. Decisive information concerning hadronic acceleration mechanisms is also likely to come from future neutrino telescopes like IceCube \citep{Ahrens2004}.

\enlargethispage*{1.0cm}

\section*{Acknowledgements}

We would like to thank the IAC for the excellent working
conditions at the Observatory de los Muchachos in La Palma. The
support of the German BMBF and MPG, the Italian INFN and the
Spanish CICYT is gratefully acknowledged. This work was also
supported by ETH Research Grant TH~34/04~3 and the Polish MNiI Grant 1P03D01028.

\end{document}